# Using Board Games and Mathematica®  
# to Teach the Fundamentals of  
# Finite Stationary Markov Chains

Roger Bilisoly[1]

[1] Department of Mathematical Sciences, Central Connecticut State University,  
1615 Stanley Ave, New Britain, CT 06050

**Abstract**  
Markov chains are an important example for a course on stochastic processes. Simple board games can be used to illustrate the fundamental concepts. For example, a looping board game (like Monopoly®) consists of all recurrent states, and a game where players win by reaching a final square (like Chutes and Ladders®) consists of all transient states except for the recurrent ending state. With the availability of computer algebra packages, these games can be analyzed. For example, the mean times in transient states and the stationary probabilities for recurrent states are easily computed. This article shows several simple board games analyzed with Mathematica®, and indicates how more complex games can be approached.

## 1. Introduction

Markov chains are an important class of probability models. If there are a finite number of states, then the theory is completely known, although solutions are impossible to do by hand unless there are only a few states, or unless the transition matrix has a simple form. However, with the advent of computer algebra systems, Markov chains with tens or hundreds of states are easily analyzable. One class of problems amenable to this type of model are board games. These also have the advantage of being familiar to most students, and they are not difficult to understand.

This paper shows how board games can be used to teach the basics of finite Markov chains and has been tested in the classroom for the graduate course Applied Stochastic Processes at Central Connecticut State University. My students find this material helpful, and I have enjoyed teaching it. This paper has the following organization. Section 2 reviews the basic theory of Markov chains, which is then applied to simple board games. Section 3 analyzes the game Monopoly, and the last section notes that there are several papers in the statistical literature that apply Markov chains to board games, which are listed under References.

## 2. Basic Theory and Examples

We start with a few definitions from Markov chain theory. First, the squares of a board game need to be matched to states of a Markov chain. Sometimes this can be done with a one-to-one mapping, though having multiple states per square can happen, too. There are two types of states: transient and recurrent. The former are visited a finite number of times almost surely, and the latter, once reached, will be revisited infinitely often almost surely. For example, the game of Chutes and Ladders has 100 squares and each square can be modeled with one state. The first 99 states are transient because a player will reach the end almost surely. The final state is recurrent since once there, the game ends, and the Markov chain remains there forever.

For actual board games, aperiodic states are typical, although changing how players move in a game can produce periodic states. For example, a Monopoly board has 40 squares, and if movement were determined by flipping a coin where heads means moving one square forward, and tails means moving one square backwards, then all the squares would be periodic with period 2. Real board games are also finite in size, so all recurrent states are positive recurrent.

Hence, looping board games such as Monopoly are ergodic, since all states are positive recurrent and aperiodic (when using two dice to move pieces).

For board games, two questions are key. First, for each transient state, what is the expected number of visits over the course of the game? For each recurrent state, what is the long-term probability of landing there? Using linear algebra, both of these questions are straightforward to answer, which is done in the next section.

## 2.1 Linear Algebra to the Rescue

Analyzing finite Markov chains requires computing the probability transition matrix, $P$. Name its entries $p_{ij}$, where this is the probability of the transition from state $i$ to $j$. The states can be reordered so that all the transient ones are first (if any), then all the recurrent ones. Then let $P_T$ be the transition matrix from transient to transient states; let $P_{TR}$ be the probabilities of going from a transient to a recurrent state; and let $P_R$ be the recurrent to recurrent probabilities. Since there is zero probability that a recurrent state can reach a transient one, we can write $P$ as follows.

$$P = \begin{pmatrix} P_T & P_{TR} \\ 0 & P_R \end{pmatrix} \tag{1}$$

For transient states, there are two fundamental questions to answer. First, if states $i$ and $j$ are both transient, then what is the expected number of times of being in state $j$ when starting from state $i$? Let us call this value $e_{ij}$ (where $e$ stands for *expectation*). Second, if state $i$ is transient and state $j$ is recurrent, what is the probability of starting in $i$ and finally reaching $j$? Call this probability $f_{ij}$ (where $f$ stands for *final probability*). Finally, let us define the following matrices: $E = (e_{ij})$ and $F = (f_{ij})$. In addition, for a game consisting of recurrent states, the goal is to find what the long-term probability of being in state $i$. Call this $\pi_i$, and call the entire vector $\pi$.

Fortunately, using linear algebra, it is easy to state the solutions to the above three questions. We merely need to compute the following equations.

$$E = (I - P_T)^{-1} \tag{2}$$

$$F = SP_{TR} \tag{3}$$

$$\pi^T P = \pi^T \tag{4}$$

Using a computer algebra software package, all three of these equations are easy to solve because matrix inversion, matrix multiplication and finding eigenvalues are all standard tasks. In fact, we see below that much of the programming is for creating the matrix $P$. Next, we apply equations (2) through (4) to some simple board games.

## 2.2 Linear Board Games

The simplest board game is linear, such as the one shown in Figure 1. Here a player starts in the left-most square, and a randomization device determines moves until the player reaches the right-most square. This can be modeled with a Markov chain with ten states, and all these except for the one corresponding to End are transient. To make calculations of the expected number of visits in each square, we need to specify how a player moves. Suppose a fair coin is flipped with the following results: move two to the right for heads, and one to the right for tails. Then the transition matrix $P$ is given in Figure 2, where a player in square 9 (from the left) is guaranteed to get to End in the next turn. Some board games require that a player must reach the end by moving the correct number of squares. If this is desired here, then replace the ninth row by a copy of the eighth row.

Figure 3 gives the result of Equation (2). The first row gives the expected number of times stopping in each square for a player beginning at Start. Note that the first entry is 1, but this must be true since the player begins at Start (which counts as one move), and then must move to the right after the first flip. Also note that the average move forward is 1.5 squares, which suggests a limiting value of $1/1.5 = 2/3$ as the number of squares increases, which is plausible in this case.

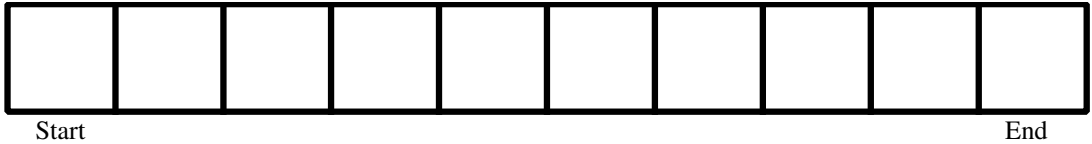

**Figure 1:** The simplest board game is linear. This one has ten squares: the first nine are transient, and the last one is recurrent.

$$P = \begin{pmatrix} 0 & \frac{1}{2} & \frac{1}{2} & 0 & 0 & 0 & 0 & 0 & 0 & 0 \\ 0 & 0 & \frac{1}{2} & \frac{1}{2} & 0 & 0 & 0 & 0 & 0 & 0 \\ 0 & 0 & 0 & \frac{1}{2} & \frac{1}{2} & 0 & 0 & 0 & 0 & 0 \\ 0 & 0 & 0 & 0 & \frac{1}{2} & \frac{1}{2} & 0 & 0 & 0 & 0 \\ 0 & 0 & 0 & 0 & 0 & \frac{1}{2} & \frac{1}{2} & 0 & 0 & 0 \\ 0 & 0 & 0 & 0 & 0 & 0 & \frac{1}{2} & \frac{1}{2} & 0 & 0 \\ 0 & 0 & 0 & 0 & 0 & 0 & 0 & \frac{1}{2} & \frac{1}{2} & 0 \\ 0 & 0 & 0 & 0 & 0 & 0 & 0 & 0 & \frac{1}{2} & \frac{1}{2} \\ 0 & 0 & 0 & 0 & 0 & 0 & 0 & 0 & 0 & 1 \\ 0 & 0 & 0 & 0 & 0 & 0 & 0 & 0 & 0 & 1 \end{pmatrix}$$

**Figure 2:** The probability transition matrix, *P,* for the board given in Figure 1 and movement is determined by flipping a coin. Heads means move two to the right, tails just one to the right.

$$E = \begin{pmatrix} 1. & 0.5 & 0.75 & 0.625 & 0.6875 & 0.65625 & 0.671875 & 0.664063 & 0.667969 \\ 0. & 1. & 0.5 & 0.75 & 0.625 & 0.6875 & 0.65625 & 0.671875 & 0.664063 \\ 0. & 0. & 1. & 0.5 & 0.75 & 0.625 & 0.6875 & 0.65625 & 0.671875 \\ 0. & 0. & 0. & 1. & 0.5 & 0.75 & 0.625 & 0.6875 & 0.65625 \\ 0. & 0. & 0. & 0. & 1. & 0.5 & 0.75 & 0.625 & 0.6875 \\ 0. & 0. & 0. & 0. & 0. & 1. & 0.5 & 0.75 & 0.625 \\ 0. & 0. & 0. & 0. & 0. & 0. & 1. & 0.5 & 0.75 \\ 0. & 0. & 0. & 0. & 0. & 0. & 0. & 1. & 0.5 \\ 0. & 0. & 0. & 0. & 0. & 0. & 0. & 0. & 1. \end{pmatrix}$$

**Figure 3:** The *E* matrix for the board given in Figure 1 and movement is determined by the transition probability matrix in Figure 2.

Figure 4 shows the Mathematica code used to create Figures 2 and 3. The most complex part of the code defines the function `tranMatrixLinear[]`, which creates the *P* matrix. This function requires the length of the linear board as well as the probabilities for moving zero, one, two, three, and so forth, squares forward. Hence the second argument consisting of {0, 0, 1}/2, means a probability of ½ of moving one or two squares forward. The matrix $P_T$ is given by `pt`, and *E* is given by `e`. Finally, the last statement prints out the matrix *E* in a matrix format.

```
tranMatrixLinear[n_,probs_]:= Module[{p={},r,i},
   Do[AppendTo[p,
     {Table[0, {i,1,r-1}], probs[[1;;Min[n-r+1, Length[probs] ] ]],
     {Table[0, {i, r + Length[probs], n}]} }//Flatten], {r,1,n}]
   Do[p[[r, n]] = 1 - Fold[Plus, 0, p[[r,1;;n-1]] ], {r,1,n}];
   Return[p]
 ]

 n = 10; (* Size of board *)
 p = tranMatrixLinear[n, {0,1,1}/2];
 pt = p[[1;;n-1, 1;;n-1]];
 e = Inverse[IdentityMatrix[n-1] - pt];
 e//MatrixForm//N
```

**Figure 4:** Code that produced the matrices shown in Figures 2 and 3. Note that performing the linear algebra tasks is easy using Mathematica.

We end this section with the gambler's ruin problem. The function `tranMatrixLinear[]` does not allow moves from right to left, so we require another one to model a possibly biased coin flip, where heads means winning a dollar and tails means losing one. The amount of the player's bankroll can be recorded by his or her position on a linear board. For example, suppose a player quits either when his or her bankroll reaches $0 or $10, where all the bets are $1 paid at even odds, although the probability of winning may differ from ½. Then the current size of the bankroll can kept track of using an 11 square linear board. A win moves one square to the right, and a loss one square to the left. A transition probability matrix is needed, and Figure 5 shows a function to create one for a board of `n+1` squares and probability of winning equal to `pWin`.

```
tranMatrixCoin[n_, pWin_] := Module[{p={}, probs},
   probs = {1-pWin, 0, pWin};
   AppendTo[p, {1, Table[0,{i,2,n}]}//Flatten];
   Do[AppendTo[p, {Table[0,{i,1,r-2}], probs, Table[0, {i, r+2, n}]}//Flatten],
     {r,2,n-1}];
   AppendTo[p, {Table[0, {i, 2, n}], 1}//Flatten];
   Return[p]
 ]
```

**Figure 5:** Code to create a probability transition matrix for the gambler's ruin problem where the game ends at $0 or $n, and the probability of winning each play is `pWin`.

```
n = 11;
p = tranMatrixCoin[n, 18/38];
pt = p[[2;;n-1,2;;n-1]];
e = Inverse[IdentityMatrix[n-2]-pt];
ptr = p[[2;;n-1,{1,n}]];
f = e.ptr; (* prob. of ending up in recurrent states *)
f//MatrixForm//N
```

**Figure 6:** Code to find the probabilities of ending up at $0 and $10 starting from $1 through $9 in the gambler's ruin problem. Here the probability of winning is 18/38, which is odds of winning in American roulette for an even money bet.

Running the code in Figure 6 produces the *F* matrix shown in Figure 7. Each line gives the odds of finishing with either $0 or $10. The first line corresponds to starting with $1, the second line starting with $2, and so forth until the ninth line. Notice that the odds are asymmetric: starting at $1 has a 94% chance of reaching $0, but starting at $9 only has an 85% chance of reaching $10. This is due to having the odds against the player. Moving on, the next section discusses looping game boards.

$$\begin{pmatrix} 0.940518 & 0.0594822 \\ 0.874426 & 0.125574 \\ 0.800992 & 0.199008 \\ 0.719397 & 0.280603 \\ 0.628737 & 0.371263 \\ 0.528003 & 0.471997 \\ 0.416077 & 0.583923 \\ 0.291715 & 0.708285 \\ 0.153534 & 0.846466 \end{pmatrix}$$

**Figure 7:** Result of the Figure 6 code. The top line gives the odds of reaching either $0 or $10 starting at $1, the second line when starting with $2, and so forth.

## 2.3 Looping Board Games

For linear board games, most of states are transient, and we want to know the mean number of visits to these squares, which is given by the matrix $E$. If there is more than one recurrent square, then we want to know the probabilities of reaching these states, which is given by the matrix $F$. However, when the game is a loop, all states are recurrent, and all are visited an infinite number of times. In this case the long term probability of landing on each square is desired, which are the eigenvalues computed by Equation (4). As in the last section, Mathematica makes it easy to obtain solutions.

One special case is particularly simple. A doubly stochastic matrix has all its rows and columns add up to 1. If the matrix $P$ is doubly stochastic, then its eigenvalues are all the same (this is easy to prove.) In general, the rows of $P$ always sum to 1, so only the columns need to be checked. However, if the movement probabilities are the same for each square, then $P$ is doubly stochastic, and so the limiting probabilities are just the inverse of the number of squares. For example, a 4 by 4 loop has 12 squares as shown in Figure 8. Move as follows: 1 square clockwise with 10% probability, 2 squares with 20%, 4 squares with 50%, and 5 squares with 30%. Since each column adds to 1, pre-multiplying $P$ by the row vector composed of all 1/12's produces the same vector; hence this is an eigenvector with the eigenvalue of 1. So in the long-term, every square has the same probability of being visited even though the probabilities at each turn are not uniform. In the next section, we examine Monopoly, which has non-uniform eigenvalues because the moves at each square are not the same.

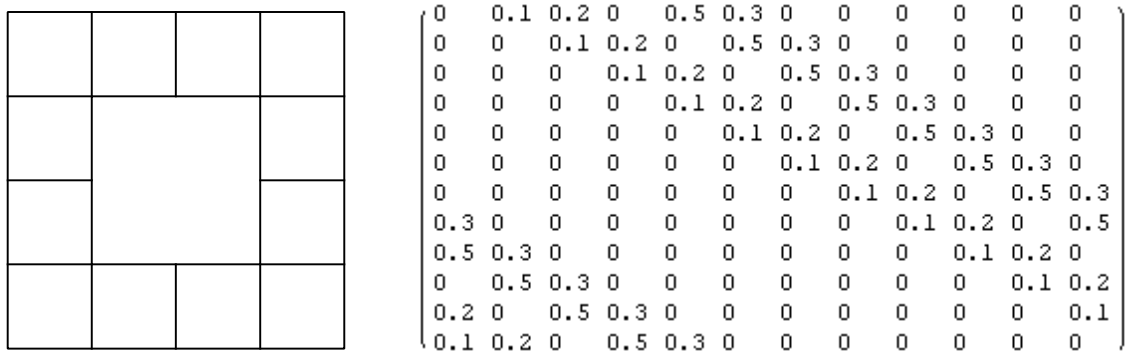

**Figure 8:** A four-by-four looping board with a doubly stochastic $P$. In the long-term, all squares have equal probability of being visited even though for each turn of the game, the probability of movement is not uniform.

## 3. An Approximate Analysis of the Game of Monopoly

Monopoly has 40 squares, and if moves were solely determined by a pair of dice, then the transition matrix *P* would be doubly stochastic, and the long-term probabilities would all be 1/40. However, there are some squares that can move a player in a manner different from the others: Chance, Community Chest and the "Go to Jail" squares. The former two require drawing a card, some of which tell the player to move. The third forces a player to go immediately to jail. We build our analysis of Monopoly starting from moves determined solely by dice, then adding the above special squares. We also assume two simplifications. First, players leave jail immediately, which is the best strategy early in the game when buying properties is important. Second, we ignore the rule that three doubles in a row forces a player to go to jail. Although this can be modeled with a Markov chain, it requires using three states per square (these states keep track of the number of doubles rolled by the player).

We start with moves solely determined by the two dice because it is useful to have this *P* in order to modify it later. The function `tranMatrixLoop[]` in Figure 9 creates this initial *P*. Here `n` is the number of squares in the loop, and `probs` is a vector containing the probabilities of moving one square forward, twos squares forward, and so forth. For Monopoly, `n` equals 40, and `probs` is the following vector: `{0,1,2,3,4,5,6,5,4,3,2,1}/36`.

```
  tranMatrixLoop[n_, probs_] := Module[{c, r, p={}, row},
    row = {probs, {Table[0, {i,Length[probs]+1, n}]}}//Flatten;
    Do[row = RotateRight[row];
      AppendTo[p,row], {i,1,n}];
    Return[p]
  ]
```

**Figure 9:** Mathematica code to create *P* for a looping board with `n` squares and movement probabilities (the same for each square) contained in the vector `probs`.

Next let us add just the "Go to Jail" square. We number the squares 1 through 40 starting at Go and going clockwise around the board. Creating the new transition probability matrix can be done by multiplying *P* by matrices, but it is not hard to do it by directly modifying it. Figure 10 shows three lines of Mathematica code that accomplishes this task. Note that square 31 is "Go to Jail" and square 11 is Jail.

```
  p = tranMatrixLoop[40, {0,1,2,3,4,5,6,5,4,3,2,1}/36];
  p[[All, 11]] += p[[All,31]];   (* Landing on 31 forces player to 11 *)
  p[[All, 31]] = Table[0,{40}]; (* There is no chance of ending a turn on 31 *)
```

**Figure 10:** Mathematica code to change the *P* found in Figure 9 so that it has a "Go to Jail" square.

The eigenvalues of this new transition matrix are given in Figure 11. Notice that square 11 (jail) now has a probability of 0.05. This is plausible because this is twice 1/40, which is the probability of the "Go to Jail" and "Jail" squares combined. All the other squares are still close to 1/40, but they are no longer equal, which is due to the non-uniform probabilities of rolling the numbers 2 through 12 with two dice. For example, the highest probability besides "Jail" is the square 7 moves clockwise from it, which reflects that two dice are most likely to roll a 7.

```
{0.0229,0.0231,0.0233,0.0236,0.0232,0.023,0.0229,0.0229,0.023,0.0231,0.05,
 0.0231,0.0239,0.0246,0.0253,0.0261,0.027,0.028,0.0276,0.0273,0.0271,0.0269,
 0.0267,0.0264,0.0268,0.027,0.0271,0.0271,0.027,0.0269,0,0.0269,0.0261,0.0254,
 0.0247,0.0239,0.023,0.022,0.0224,0.0227}
```

**Figure 11:** Square probabilities when only the "Go to Jail" square is added to the 40 square loop. Note that the probabilities are no longer uniform.

To include Chance and Community Chest, we assume that each time a card is drawn, it is replaced and then all the cards are shuffled. This makes the probability of picking any particular card equal to 1/16 (there are 16 cards each for both types.) Then the p found above can be modified so that the probabilities of reaching each destination are included. Figure 12 gives the code for the first Community Chest square. Once this is done for the other two as well as the three Chance squares, then the probability transition matrix is finished (up to the approximations noted above), and the eigenvalues can be found. Finally, note that the Chance on square 37 must be modeled before the Community Chest on 34, because the former has a card that makes the player move backwards three squares.

```
p[[All,1]]  += p[[All,3]]*1/16; (* Advance to Go *)
p[[All,11]] += p[[All,3]]*1/16; (* Advance to Jail *)
p[[All,3]]  *= 14/16;
```

**Figure 12:** Mathematica code that changes *P* so that the first Community Chest square is properly modeled. Note that only 2 cards require relocating and the rest require other actions, for example, paying or collecting money.

Figure 13 gives the eigenvalues, and these are plotted in Figure 14 using a grey scale that enhances the probability differences. Note that Illinois Ave. has the highest probability (besides Jail), which is a well-known fact among Monopoly players. Besides the "Go to Jail" square, the other three dark squares represent Chance. This arises from the high number of cards that tell the player to relocate (9 out of 16).

```
{0.03114,0.02152,0.019,0.02186,0.02351,0.02993,0.02285,0.00876,0.02347,
 0.02331,0.05896,0.02736,0.02627,0.02386,0.02467,0.02919,0.02777,0.02572,
 0.02917,0.03071,0.02875,0.0283,0.01048,0.02739,0.03188,0.03064,0.02707,
 0.02679,0.02811,0.02591,0,0.02687,0.02634,0.02377,0.0251,0.02446,0.00872,
 0.02202,0.02193,0.02647}
```

**Figure 13:** Probabilities when Chance, Community Chest and the "Go to Jail" squares are modeled by *P*. Note that the probabilities are even less uniform than those in Figure 11.

## 4. Conclusions

The above examples (and others similar to them) have been used in my teaching. I enjoy discussing board games and have received positive feedback from my students. Moreover, applying Markov chains to board games has been done several times in the literature, and the instances known to me are given in the References below. Most of these analyze either Monopoly or Chutes and Ladders. These articles can be assigned to be read and then discussed, or they can be the basis of homework problems or student projects. Since board games are easily comprehensible, and with Mathematica, easily analyzable, these are a great source of examples for teaching. Let the games begin!

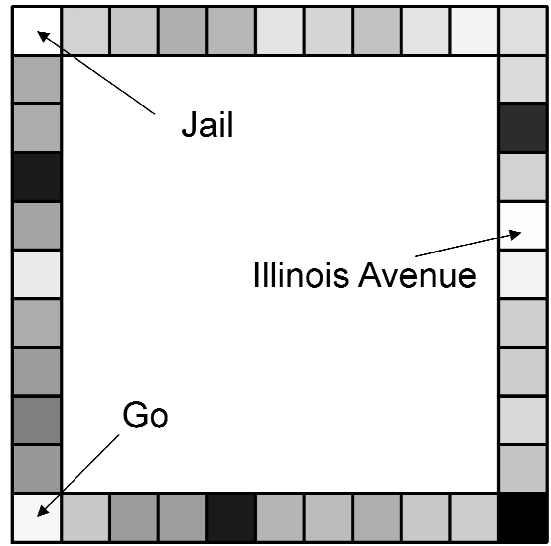

**Figure 14:** A plot of the eigenvalues given in Figure 13 using a grey scale that enhances differences among the probabilities.

Monopoly® is a registered trademark of Parker Brothers, a subsidiary of Hasbro. Chutes and Ladders® is a registered trademark of Milton Bradley, a subsidiary of Hasbro. Mathematica® is a registered trademark of Wolfram Research.